\documentstyle[12pt,epsfig]{article}
\def\be{\begin{equation}} \def\ee{\end{equation}} \def\bea{\begin{eqnarray}}
\def\eea{\end{eqnarray}} \def\nnb{\nonumber}

\begin{document}

\hfill{August 7, 2020} 

\begin{center}
\vskip 6mm 
\noindent
{\Large\bf  
Elastic $\alpha$-$^{12}$C scattering at low energies with the
sharp resonant $0_3^+$ state of $^{16}$O
}
\vskip 6mm 

\noindent
{\large 
Shung-Ichi Ando\footnote{mailto:sando@sunmoon.ac.kr}, 
\vskip 6mm
\noindent
{\it
School of Mechanical and ICT convergence engineering, \\
Sunmoon University,
Asan, Chungnam 31460,
Republic of Korea
}
}
\end{center}

\vskip 6mm

An inclusion of sharp resonant $0_3^+$ state of $^{16}$O 
and first excited $2_1^+$ state of $^{12}$C in a study 
of $s$-wave elastic $\alpha$-$^{12}$C scattering 
at low energies is investigated in an effective Lagrangian approach. 
The elastic scattering amplitude is parted into two; 
one is for the sharp resonant $0_3^+$ state of $^{16}$O parameterized
by Breit-Wigner formula, and the other is for a non-resonant
part of the amplitude parameterized by effective range expansion.
In the non-resonant part of the amplitude,
a contribution from the $2_1^+$ state of $^{12}$C is included. 
We discuss a large correlation 
between a coupling for the $2_1^+$ state of $^{12}$C and an
effective range parameter $Q_0$ 
as well as a necessity of inclusion of a vertex correction 
for the initial and final $\alpha$-$^{12}$C states.
After fixing parameters appearing in the amplitudes
by using experimental data, 
we calculate asymptotic normalization coefficients 
for ground $0_1^+$ state and first excited $0_2^+$ state 
of $^{16}$O and compare them to previous results found in literature.  

\vskip 5mm 
\noindent PACS(s): 
11.10.Ef, 
24.10.-i, 
25.55.-e, 
26.20.Fj  

\newpage
\vskip 2mm \noindent
{\bf 1. Introduction}

The radiative $\alpha$ capture on $^{12}$C,
$^{12}$C($\alpha$,$\gamma$)$^{16}$O, is a key reaction to 
determine the ratio of $^{12}$C/$^{16}$O producing in stars~\cite{f-rmp84}.
Due to subthreshold $l_{i-th}^\pi = 1_1^-$ and $2_1^+$ states of $^{16}$O
just below $\alpha$-$^{12}$C breakup threshold, the radiative $\alpha$ capture
reaction will be $E1$ and $E2$ transitions dominant while a small 
contribution comes out of so-called cascade transitions
where $\alpha$ and $^{12}$C first form an excited bound state of $^{16}$O
and it subsequently decays down to ground $0_1^+$ state of $^{16}$O. 
Asymptotic normalization coefficients (ANCs) of the bound states of $^{16}$O
play an important role to estimate the radiative $\alpha$ capture rates,
equivalently astrophysical $S$-factor at Gamow-Peak energy, $E_G=0.3$~MeV,
for $R$-matrix analysis~\cite{lt-rmp58}.
Values of the ANCs for the subthreshold $1_1^-$ and $2_1^+$ states
of $^{16}$O are converging in both theory and experiment
while scattered values 
for $3_1^-$, $0_2^+$,
and $0_1^+$ states of $^{16}$O are found in literature. 
During the last half century, numerous experimental and theoretical
studies related to the radiative $\alpha$ capture reaction have been
carried out. For review, see, e.g., 
Refs.~\cite{bb-npa06,chk-epja15,bk-ppnp16,detal-rmp17}
and references therein. 

In our recent works, we constructed an effective field theory (EFT)
for the $^{12}$C($\alpha$,$\gamma$)$^{16}$O reaction 
at $E_G$~\cite{a-epja16},
parameters appearing in an effective Lagrangian were fitted to 
experimental data for elastic $\alpha$-$^{12}$C 
scattering~\cite{a-prc18,a-jkps18,ya-jkps19} 
and $S_{E1}$ factor of the $^{12}$C($\alpha$,$\gamma$)$^{16}$O reaction 
through the $E1$ transition~\cite{a-prc19}, 
and a value of the $S_{E1}$ factor at $E_G$
was estimated in the theory for the first time~\cite{a-prc19}.\footnote{
Recently, we apply the EFT to a study for $\beta$ delayed $\alpha$ 
emission from $^{16}$N~\cite{a-19}.
}
An EFT may provide us a model independent method for theoretical calculation
at low energies, in which one needs to introduce a separation (momentum)
scale between relevant physical degrees of freedom at low energy and 
irrelevant degrees of freedom at high energy. An effective Lagrangian is
constructed using the relevant low-energy degrees of freedom and 
expanded in terms of the number of derivatives order by order.
The irrelevant degrees of freedom are integrated out of an effective 
Lagrangian, and effects from those at high energy are presumed to be 
embedded in coefficients of terms appearing in an effective Lagrangian.
Those coefficients are theoretically obtained from its mother theory 
while they are practically fixed by using experimental data.
The derivative expansion scheme provides us a perturbative expansion, 
which is useful to estimate a theoretical error for a reaction in 
question.
For review for EFTs, see, e.g., 
Refs.~\cite{bv-arnps02,bh-pr06,m-15,hjp-17,hkvk-19}.
In our previous works, we incorporate broad resonant $1_2^-$ and $3_2^-$
states of $^{16}$O in the reaction amplitudes using effective range expansion 
but could not include sharp resonant $0_3^+$ and $2_2^+$
states of $^{16}$O~\cite{ya-jkps19,a-prc19}.
In this work, we study an inclusion of a sharp resonant $0_3^+$ 
state of $^{16}$O, along with a first excited $2_1^+$ state of $^{12}$C,
in the elastic $\alpha$-$^{12}$C scattering for $l=0$ channel.  

Counting rules for a resonant state, for which one includes the Coulomb 
interaction between two charged particles, are discussed by Higa, Hammer,
and van Kolck for a halo-like system~\cite{hhvk-npa08}, 
and by Gelman for a resonance state in a Breit-Wigner form~\cite{g-prc09}.
We follow a prescription suggested by Higa, Hammer, and van Kolck to 
rewrite a scattering amplitude presented in terms of effective range
parameters to an amplitude presented by using a Breit-Wigner formula.
We also follow another prescription by Gelman to separate a scattering 
amplitude into two parts; one is an amplitude for a sharp resonant $0_3^+$ 
state of $^{16}$O, 
and the other is that for the rest of a non-resonant part of the amplitude.
As we discussed in our previous work~\cite{a-prc18},
because of a modification of the counting rules 
for the elastic $\alpha$-$^{12}$C scattering at low energies, we include 
the terms up to $p^6$ order in the effective range expansion, 
where $p$ is the magnitude of relative momentum between $\alpha$ and $^{12}$C.
In this work, parameters appearing in the amplitudes are fixed by 
using the following experimental data; 
binding energies for bound states and a resonant energy and a width for 
a resonant state of $^{16}$O and phase shift data
for elastic $\alpha$-$^{12}$C scattering 
reported by Tischhauser \textit{et al.}~\cite{tetal-prc09}.
Because the energy range for the experimental data reported by
Tischhauser \textit{et al.} is 
$2.6 \le E_\alpha \le 6.62$~MeV, where $E_\alpha$ 
is the $\alpha$ energy in laboratory frame~\footnote{
One has a relation $E_\alpha = \frac43 E$ where $E$ is the total kinetic energy
of $\alpha$ and $^{12}$C in center of mass frame.  
}, 
we include a first excited $2_1^+$ state of $^{12}$C whose excited energy 
is $E(2_1^+) = 4.44$~MeV in the present study.
We also include
a vertex correction for initial and final state interactions
between $\alpha$ and $^{12}$C in a phenomenological way. 

Two parameters appearing in the scattering amplitude for the sharp resonance
state are fixed by experimental values for resonance energy and width for
the $0_3^+$ state of $^{16}$O while two effective parameters appearing
in the non-resonant part of the amplitude are fixed by using binding energies
of ground $0_1^+$ state and first excited $0_2^+$ state of $^{16}$O. 
For other four parameters,
$P_0$ and $Q_0$ for the effective range parameters,
a coupling constant $\tilde g_0$ for a contribution from the $2_1^+$ state
of $^{12}$C, and a constant $R_2$ for a vertex correction for the initial
and final states of $\alpha$ and $^{12}$C, are fitted to experimental
phase shift data $\delta_0$ for the elastic $\alpha$-$^{12}$C scattering
for $l=0$ channel. 
We find that the experimental data including the sharp resonant $0_3^+$ state
of $^{16}$O at the energy range, $2.6  \le E_\alpha \le 6.62$~MeV, reported
by Tischhauser \textit{et al.}
are well reproduced while a coupling $\tilde g_0$ is not 
fixed by using the phase shift data because of a large correlation with $Q_0$.
We then calculate asymptotic normalization coefficients (ANCs) for the 
$0_1^+$ and $0_2^+$ states of $^{16}$O and
compare our results with those found in previous studies. 

The present article is organized as follows.
In section 2, an effective Lagrangian for elastic $\alpha$-$^{12}$C scattering
for $l=0$ channel including a sharp resonant $0_3^+$ state of $^{16}$O
and a first excited $2_1^+$ state of $^{12}$C are displayed. In section 3,
scattering amplitudes for the part of the sharp resonance $0_3^+$ state 
of $^{16}$O
and for the rest of the non-resonant part of the amplitude are derived.
In section 4, four parameters are fixed by using the two binding energies 
and a resonant energy and a width of the $0_1^+$, $0_2^+$, and 
$0_3^+$ states of $^{16}$O and remaining parameters
are fitted to the experimental phase shift data for the elastic 
$\alpha$-$^{12}$C scattering for $l=0$ channel. 
We then calculate ANCs for the $0_1^+$ and $0_2^+$ states of $^{16}$O
and compare our results to those found in previous works. 
In section 5, results and discussion of the present work are presented.

\vskip 2mm \noindent
{\bf 2. Effective Lagrangian}

An effective Lagrangian to derive a scattering amplitude
for a study of $s$-wave elastic
$\alpha$-$^{12}$C scattering at low energies including a sharp
resonant $0_3^+$ state of $^{16}$C and a first excited $2_1^+$ state
of $^{12}$C may be written 
as~\cite{a-epja16,a-prc18,a-jkps18,a-epja07}
\bea
{\cal L} &=& \phi_\alpha^\dagger \left(
iD_0 
+\frac{\vec{D}^2}{2m_\alpha}
+ \cdots
\right) \phi_\alpha
+ \phi_C^\dagger\left(
iD_0
+ \frac{\vec{D}^2}{2m_C}
+\cdots
\right)\phi_C
\nnb \\ &&
+
\phi_{C,ij}^{(l=2)\dagger} \left(
i D_0
+ \frac{1}{2m_C}\vec{D}^2
- \Delta_{(2)}
+ \cdots
\right)\phi_{C,ij}^{(l=2)}
\nnb \\ && +
\sum_{n=0}^3 
C_{n}^{(rs)}d_{(rs)}^\dagger 
\left[
iD_0 
+ \frac{\vec{D}^2}{2(m_\alpha+m_C)}
\right]^n d_{(rs)}
- y_{(rs)}\left[
d_{(rs)}^\dagger(\phi_\alpha  \phi_C)
+ (\phi_\alpha  \phi_C)^\dagger d_{(rs)}
\right] 
\nnb \\ && +
\sum_{n=0}^3 
C_{n}^{(nr)}d_{(nr)}^\dagger 
\left[
iD_0 
+ \frac{\vec{D}^2}{2(m_\alpha+m_C)}
\right]^n d_{(nr)}
- y_{(nr)}\left[
d_{(nr)}^\dagger(\phi_\alpha  \phi_C)
+ (\phi_\alpha  \phi_C)^\dagger d_{(nr)}
\right] 
\nnb \\ && 
- y_{(rs)}' \left[
d_{(rs)}^\dagger \left(
\phi_\alpha O_0' \phi_C
\right) + \left(
\phi_\alpha O_0' \phi_C
\right)^\dagger d_{(rs)}
\right]
- y_{(nr)}' \left[
d_{(nr)}^\dagger \left(
\phi_\alpha O_0' \phi_C
\right) + \left(
\phi_\alpha O_0' \phi_C
\right)^\dagger d_{(nr)}
\right]
\nnb \\ &&
-g_0 \left[
d^\dagger_{(nr)}\left(
\phi_\alpha O_{2,ij}\phi_{C,ij}^{(l=2)}
\right) +\left(
\phi_\alpha O_{2,ij}\phi_{C,ij}^{(l=2)}
\right)^\dagger d_{(nr)}
\right]
+ \cdots
\label{eq;Lagrangian}
\,,
\eea
where $\phi_\alpha$ ($m_\alpha$) and 
$\phi_C$ ($m_C$) are scalar fields (masses) of $\alpha$ and $^{12}$C, 
respectively.  
$D^\mu$ is a covariant derivative,
$D^\mu = \partial^\mu + i {\cal Q}A^\mu$ where
${\cal Q}$ is a charge operator and $A^\mu$ is the photon field.
The dots denote higher-order terms. 
$\phi_{C,ij}^{(l=2)}$ ($\Delta_2$) is a field 
(an excited energy) for the first excited $2_1^+$ state of 
$^{12}$C. 
$d_{(rs)}$ and $d_{(nr)}$ are composite fields of $^{16}$O consisting of 
$\alpha$ and $^{12}$C fields for $l=0$ channel for the sharp resonant (rs) 
and the non-resonant (nr) parts, respectively, which are introduced 
for perturbative expansion around the unitary 
limit~\cite{b-pr49,k-npb97,bs-npa01,ah-prc05}.  
The coupling constants, 
$C_n^{(rs)}$ and $C_n^{(nr)}$ with $n=0,1,2$, and 3, correspond to 
the effective range parameters of elastic $\alpha$-$^{12}$C
scattering while
the coupling constants $y_{(rs)}$ and $y_{(nr)}$ are redundant
\footnote{
In the denominator of the elastic scattering amplitudes, 
the couplings appear in the form, $C_n^{(rs,nr)}/y_{(rs,nr)}^2$ 
with $n=0,1,2,3$, and are fitted to, e.g., the effective range parameters, 
$1/a_0$, $r_0$, $P_0$, $Q_0$, for $l=0$, respectively. 
The $y_{(rs,nr)}$ couplings are redundant, 
and one can arbitrarily fix their values.
},
which are conventionally taken 
as $y_{(rs)} = y_{(nr)} =\sqrt{2\pi/\mu}$.
$y_{(rs)}'$ and $y_{(nr)}'$ are higher-order vertex corrections for 
$d_{(rs,nr)}$-$\alpha$-$^{12}$C vertices at next-to-next-to leading
order (NNLO).
We note that we will not mix these two fields, 
$d_{(rs)}$ and $d_{(nr)}$, for the sharp resonant part and 
the non-resonant part of the amplitudes 
through the $y_{(rs)}$ and $y_{(nr)}$ interactions or 
the $y_{(rs)}'$ and $y_{(nr)}'$ interactions.
In addition, we include the $y_{(rs)}'$ and $y_{(nr)}'$
interactions only in the initial and final states of $\alpha$ and $^{12}$C.
We will discuss those issues later. 
$g_0$ is a coefficient for the transitions 
between $s$-wave $\alpha$ and first excited $2_1^+$ state of $^{12}$C and
the non-resonant part of the composite $^{16}$O field, $d_{(nr)}$.
The operators are given as 
\bea
&& 
O_{2,ij} =
- \frac{\stackrel{\leftrightarrow}{D}_i}{M}
 \frac{\stackrel{\leftrightarrow}{D}_j}{M}
+ \frac13 \delta_{ij}
\frac{\stackrel{\leftrightarrow}{D}^2}{M^2}
\,,
\ \ \ 
O_0' = - \frac{\stackrel{\leftrightarrow}{D}^2}{M^2}
\,,
\ \ \ 
i\frac{\stackrel{\leftrightarrow}{D}_i}{M}
\equiv
i \left(
\frac{\stackrel{\rightarrow}{D}_C}{m_C} -
\frac{\stackrel{\leftarrow}{D}_\alpha}{m_\alpha} 
\right)_i
\,.
\eea

\vskip 2mm \noindent
{\bf 3. Amplitudes for the elastic scattering}

The elastic scattering amplitude $A_0$ is decomposed into two parts:
\bea
A_0 &=& A_0^{(rs)} + A_0^{(nr)}\,,
\label{eq;A0}
\eea
where 
$A^{(rs)}$ is the amplitude for the sharp resonant $0_3^+$ state of $^{16}$O
parameterized by the Breit-Wigner formula and  
$A_0^{(nr)}$ is the scattering amplitude for the non-resonant
contribution parameterized by the effective range expansion.

\begin{figure}[t]
\begin{center}
\resizebox{0.7\textwidth}{!}{
  \includegraphics{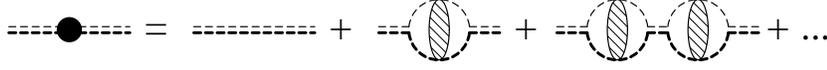}
}
\caption{
Diagrams for dressed $^{16}$O propagator.
A thick (thin) dashed line represents a propagator of $^{12}$C ($\alpha$),
and a thick and thin double dashed line with and without a filled circle
represent a dressed and bare $^{16}$O propagator, respectively.
A shaded blob represents
a set of diagrams consisting of all possible one-potential-photon-exchange
diagrams up to infinite order and no potential-photon-exchange one.
}
\label{fig;propagator}       
\end{center}
\end{figure}
\begin{figure}[t]
\begin{center}
\resizebox{0.2\textwidth}{!}{%
  \includegraphics{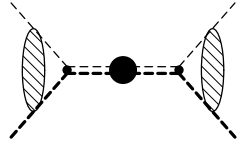}
}
\caption{
Diagram of the scattering amplitude.
See the caption of Fig.~\ref{fig;propagator} as well.
}
\label{fig;scattering_amplitude}      
\end{center}
\end{figure}

Both the scattering amplitudes $A_0^{(rs)}$ and $A_0^{(nr)}$ 
are calculated from the diagrams depicted in 
Figs.~\ref{fig;propagator} and \ref{fig;scattering_amplitude}.
For the elastic scattering amplitude for the sharp resonant $0_3^+$ state
of $^{16}$O, we may first write it 
in terms of the effective range expansion as
\bea
A^{(rs)}_0 &=&
\frac{2\pi}{\mu}
\frac{e^{2i\sigma_0}C_\eta^2F(p)^2}{
K^r_0(p)
-2\kappa H(\eta)
}
\label{eq;A0_rs}
\,,
\eea
with
\bea
e^{2i\sigma_0} &=& \frac{\Gamma(1+i\eta)}{\Gamma(1-i\eta)}\,,
\ \ \ 
C_\eta^2 =  \frac{2\pi\eta}{e^{2\pi\eta}-1}\,,
\ \ \
H(\eta) = \psi(i\eta) + \frac{1}{2i\eta} -\ln(i\eta)\,,
\eea
where $\psi(z)$ is the digamma function. 
As mentioned before,
$p$ is the magnitude of relative momentum between $\alpha$ and
$^{12}$C and $\eta = \kappa/p$ where $\kappa$ is the inverse of the Bohr
radius,
$\kappa = Z_\alpha Z_C\mu\alpha_E$: 
$Z_\alpha$ and $Z_C$ are the numbers of protons in $\alpha$ and $^{12}$C, 
$\mu$ is the reduced mass of $\alpha$ and $^{12}$C,
and $\alpha_E$ is the fine-structure constant. 
We note that the Coulomb self-energy term, $-2\kappa H(\eta)$
is obtained from a bubble diagram due to propagation of the ground states 
of $\alpha$ and $^{12}$C.
Here we have ignored the self-energy contribution from the $2_1^+$ state 
of $^{12}$C because the amplitude will be 
rewritten as the Breit-Wigner-like expression below, which has a sharp 
peak at the resonant energy $E_r$, and the off-peak energy contribution 
will be regarded as a higher order correction.  
The functions $F(p)$ and $K^r_0(p)$ contain dynamics
for the elastic scattering through the sharp resonant state.
The function $F(p)$ is a vertex correction of the initial and final 
state interactions between $\alpha$ and $^{12}$C while the function $K^r_0(p)$ 
is a polynomial function expanded around the unitary limit, which is
presented in terms of the effective range parameters. Thus, one has
\bea
F(p) &=& 1-\frac16R_2p^2\,,
\label{eq;Fr}
\\
K^r_0(p) &=&
-\frac{1}{a_0^r}
+\frac12 r_0^r p^2
-\frac14 P_0^r  p^4
+Q_0^r p^6\,,
\label{eq;K0r}
\eea
where we have introduced a squared radius like parameter $R_2$,
$R_2 = -6y_{(rs)}'/(y_{(rs)}\mu^2)$ in Eq.~(\ref{eq;Fr}), 
and the coefficients,
$C_n^{(rs)}/y_{(rs)}^2$ with $n=0,1,2,3$, are replaced by 
the effective range parameters in Eq.~(\ref{eq;K0r}).  

Following a prescription 
suggested by Higa, Hammer, and van Kolck~\cite{hhvk-npa08}
to rewrite the amplitude parameterized by 
the effective range expansion to that by the Breit-Wigner formula,
we have 
\bea
A_0^{(rs)} &=&
- \frac{2\pi}{\mu}
\frac{e^{2i\sigma_0}}{\sqrt{2\mu E}}
\frac{\frac12\Gamma(E) F(p)^2}{Z_rD^r(E) + i\frac12\Gamma(E)}\,,
\eea
with
\bea
\Gamma(E) &=& \Gamma_r\frac{
e^{2\pi\eta_r}-1
}{
e^{2\pi\eta}-1
}\,,
\\
Z_rD^r(E) &=& E- E_r +Z_r\left\{
\frac{}{}
\mu^2P_0^r(E-E_r)^2
-8\mu^3Q_0^r(E+2E_r)(E-E_r)^2
\right. \nnb \\ && \left.
+2\kappa\left[
ReH(\eta) -ReH(\eta_r)
- \left. \frac{\partial}{\partial E} Re H(\eta)\right|_{E=E_r}(E-E_r)
\right]
\right\}\,,
\label{eq;ZD}
\\
Z_r^{-1} &=&
\left.
\frac{\partial}{\partial E} D^r(E)\right|_{E=E_r}\,,
\ \ \
Z_r = \frac{e^{2\pi\eta_r}-1}{4\pi\kappa}
\Gamma_r\,,
\eea
where $E$ is the energy of the $\alpha$-$^{12}$C system in the 
center of mass frame, $E=p^2/(2\mu)$, and  
$E_r$ and $\Gamma_r$ are the energy and the width of the
resonant $0_3^+$ state of $^{16}$O, which are related to two effective 
range parameters, $a_0^r$ and $r_0^r$ in Eq.~(\ref{eq;K0r}).
$P_0^r$ can be fixed by using a condition that a large contribution
from the Coulomb self-energy term to the $p^4$ term is cancelled 
with the $P_0^r$ term. Thus, we have 
\bea
P_0^r &=&
24\mu E_r Q_0^r
- \left. \frac{\kappa}{\mu^2}
\frac{d^2}{dE^2}ReH(\eta)
\right|_{E=E_r}\,,
\eea
where $Q_0^r$ can be chosen arbitrarily.
Thus, the scattering amplitude $A_0^{(rs)}$ for the sharp resonant state
is represented by the three parameters, $R_2$, $E_r$, and $\Gamma_r$. 

For the non-resonant amplitude $A_0^{(nr)}$, we have 
\bea
A^{(nr)}_0 &=&
\frac{2\pi}{\mu}
\frac{e^{2i\sigma_0}C_\eta^2F(p)^2}{
K_0(p)
-2\kappa \left[
H(\eta)
+ \frac{2\tilde g_0^2}{3\mu^4}
H_2(\tilde\eta)
\right] }
\label{eq;Aer0}
\,,
\eea
with $\tilde{g_0} = g_0/y_{(nr)}$ and
\bea
H_2(\tilde\eta) &=& W_2(\tilde{p}) H(\tilde{\eta})\,,
\ \ \
W_2(\tilde{p}) = \frac14(\kappa^2+\tilde{p}^2)(\kappa^2+4\tilde{p}^2)\,,
\\
\tilde\eta &=& \kappa/\tilde{p}\,,
\ \ \
\tilde{p} = i\sqrt{-2\mu(E-\Delta_2)-i\epsilon}\,,
\eea
where $\Delta_2$ is an excitation energy of the $2_1^+$ state of $^{12}$C,
$\Delta_2=4.440$~MeV. 
The second Coulomb self-energy term, $-2\kappa H_2(\tilde\eta)$, is 
obtained from a bubble diagram propagating the ground state $\alpha$ and 
the excited $2_1^+$ state of $^{12}$C 
where those two states
are in relative $d$-wave state and coupled to the $s$-wave 
composite $^{16}$O field for the non-resonant contribution. 
Interaction between $\alpha$ and $^{12}$C is parameterized
in the function $K_0(p)$ by the effective range expansion;
one has
\bea
K_0(p) &=&
-\frac{1}{a_0}
+\frac12 r_0p^2
-\frac14 P_0 p^4
+Q_0 p^6\,.
\eea

We fix two parameters among the four effective range parameters,
$a_0$, $r_0$, $P_0$, and $Q_0$, by using conditions that 
the inverse of the scattering amplitude $A_0^{(nr)}$ vanishes at 
the energies of the ground $0_1^+$ state and the first excited $0_2^+$
state of $^{16}$O. Thus, the denominator of the scattering amplitude
vanishes,
\bea 
D_0(p) = K_0(p) -2\kappa\left[ H(\eta) 
+ \frac{2\tilde{g}_0^2}{3\mu^4}H_2(\tilde{\eta})
\right] =0\,,  
\label{eq;binding_poles}
\eea
at $p=i\gamma_0$ and $p=i\gamma_1$ where $\gamma_0$ and $\gamma_1$ are
binding momenta for the $0_1^+$ and $0_2^+$ states of $^{16}$O, respectively; 
$\gamma_{0,1} = \sqrt{2\mu B_{0,1}}$ where $B_0$ and $B_1$
are the binding energies for the $0_1^+$ and $0_2^+$ states of $^{16}$O
from the $\alpha$-$^{12}$C breakup threshold, respectively.   
Using the conditions from Eq.~(\ref{eq;binding_poles}), 
we fix two effective range parameters, 
$a_0$ and $r_0$ as
\bea
\frac{1}{a_0} &=& 
\frac14\gamma_0^2\gamma_1^2 P_0
+(\gamma_0^4\gamma_1^2+\gamma_0^2\gamma_1^4)Q_0 
\nnb \\ && 
+\frac{2\kappa}{\gamma_0^2-\gamma_1^2}\left\{
\gamma_1^2 \left[
H(\eta_{b0}) + \frac{2\tilde{g}_0^2}{3\mu^4}H_2(\tilde\eta_{b0})
\right]
-\gamma_0^2 
\left[
H(\eta_{b1}) + \frac{2\tilde{g}_0^2}{3\mu^4}H_2(\tilde\eta_{b1})
\right]
\right\}\,,
\label{eq;a0}
\\
r_0 &=& 
-\frac12(\gamma_0^2+\gamma_1^2)P_0 
-2(\gamma_0^4+\gamma_0^2\gamma_1^2+\gamma_1^4)Q_0
\nnb \\ &&
-\frac{4\kappa}{\gamma_0^2-\gamma_1^2}\left\{
\left[
H(\eta_{b0}) + \frac{2\tilde{g}_0^2}{3\mu^4}H_2(\tilde\eta_{b0})
\right]
- \left[
H(\eta_{b1}) + \frac{2\tilde{g}_0^2}{3\mu^4}H_2(\tilde\eta_{b1})
\right]
\right\}\,,
\label{eq;r0}
\eea
where $\eta_{b0,b1} = \kappa/(i\gamma_{0,1})$
and $\tilde\eta_{b0,b1} = \kappa/(i\sqrt{\gamma_{0,1}^2+2\mu \Delta_2})$.
Using the two relations in Eqs.~(\ref{eq;a0}) and (\ref{eq;r0}), 
we rewrite the denominator of the amplitude $D_0(p)$ as
\bea
D_0(p) &=& 
-\frac14\left[
\gamma_0^2\gamma_1^2 + (\gamma_0^2+\gamma_1^2)p^2 + p^4
\right] P_0 
\nnb \\ && 
+\left[
-\gamma_0^4\gamma_1^2 -\gamma_0^2\gamma_1^4
-(\gamma_0^4 +\gamma_0^2\gamma_1^2 +\gamma_1^4) p^2 
+ p^6
\right]Q_0 
\nnb \\ && -2\kappa\left\{
\frac{\gamma_1^2+p^2}{\gamma_0^2-\gamma_1^2}\left[
H(\eta_{b0}) + \frac{2\tilde{g}_0^2}{3\mu^4}H_2(\tilde{\eta}_{b0})
\right]
-\frac{\gamma_0^2+p^2}{\gamma_0^2-\gamma_1^2}\left[
H(\eta_{b1}) + \frac{2\tilde{g}_0^2}{3\mu^4}H_2(\tilde{\eta}_{b1})
\right]
\right. \nnb \\ && \left. \frac{}{}
+H(\eta) + \frac{2\tilde{g}_0^2}{3\mu^4}H_2(\tilde{\eta})
\right\}\,,
\eea
where we have three constants, $P_0$, $Q_0$, $\tilde{g}_0$
in the function $D_0(p)$ and one constant $R_2$ in the function $F(p)$
for the non-resonant amplitude $A_0^{(nr)}$ to fix by using the 
phase shift data. We note that we use the same parameter $R_2$ for 
both the amplitudes $A_0^{(rs)}$ and $A_0^{(nr)}$ because the parameter
$R_2$ commonly appears in the initial and final state interactions
between $\alpha$ and $^{12}$C.     
Thus, we have six parameters $\{P_0,Q_0,\tilde g_0,R_2,E_r,\Gamma_r\}$
in the scattering amplitude $A_0$. 

The ANCs $|C_b|_0$ and $|C_b|_1$ for the $0_1^+$ and $0_2^+$ states
of $^{16}$O are calculated by using a formula
\bea
|C_b|_n &=& \Gamma(1+\eta_{bn})F(i\gamma_{n})
\left[
\left.
(-1)^n \frac{\partial D_0(p)}{\partial p^2}
\right|_{p^2 = -\gamma_{n}^2}
\right]^{-1/2}\,,
\label{eq;Cb}
\eea
with $n=0$ or $n=1$, where we have included the vertex correction $F(p)$ 
in the expression for the ANCs found in Ref.~\cite{ir-prc84}.

\vskip 2mm \noindent
{\bf 4. Numerical results}

To fix the coefficients appearing in the scattering amplitude $A_0$ 
for $l=0$ channel,
we employ the data for the phase shift $\delta_0$ reported 
by Tischhauser \textit{et al.}\cite{tetal-prc09}.
The elastic scattering amplitude for $l=0$ in terms of the phase 
shift $\delta_0$ is given as
\bea\nnb
A_0 &=&
\frac{2\pi}{\mu}\frac{e^{2i\sigma_0}}{p\cot\delta_0-ip}\,.
\eea
Because we represent the scattering amplitude $A_0$ as 
two terms, $A_0^{(rs)}$ and $A_0^{(nr)}$,
as given in Eq.~(\ref{eq;A0}), 
we fit the parameters to the data 
by using a relation for the squared amplitude as
\bea\nnb
\frac{1}{p^2}\sin^2\delta_0
&=&
\left|
\frac{1}{p\cot\delta_0-ip}
\right|^2
= \frac{\mu^2}{4\pi^2}\left|
A_0^{(rs)} + A_0^{(nr)}
\right|^2\,.
\eea

As mentioned above, the six parameters 
$\{P_0,Q_0,\tilde g_0,R_2,E_r,\Gamma_r\}$ remain in $A_0$ while
the resonant energy $E_r$ and its width $\Gamma_r$ for the 
sharp resonant $0_3^+$ state of $^{16}$O are experimentally known 
as $E_r = 4.887(2)$~MeV and $\Gamma_r=1.5(5)$~keV~\cite{twc-npa93}.
(The corresponding laboratory energy of $E_r$
is $E_{\alpha,r} = \frac43 E_r = 6.516(3)$~MeV.)
We include them in the fitting because more precise adjustment
is necessary to reproduce the sharp peak appearing in the data. 
$R_2$ basically accounts for the slowly varying shape of the phase shift 
at high-energy region, $5.5 \le E_\alpha \le 6.62$~MeV. 
$\tilde g_0$ is a dimensionless parameter and represents a contribution
from the first excited $2_1^+$ state of $^{12}$C.  
As will see below, we find that the parameter $\tilde g_0$ is strongly
correlated to $Q_0$ and cannot be determined from the phase shift data.  
Because we have no restriction for $\tilde g_0$, 
we give some values for $\tilde g_0$ 
(here we arbitrarily choose $\tilde g_0 = 0, 20, 40$, and 60) 
and fit the remaining five parameters 
$\{P_0,Q_0,R_2,E_r,\Gamma_r\}$ to the phase shift data
employing a standard $\chi^2$ fit~\footnote{
We employ a python package, {\tt emcee}\cite{fmetal-12},
for the fitting.
\label{footnote;emcee}
}.
As mentioned above,
a reported energy range of the data is $2.6 \le E_\alpha \le 6.62$~MeV where
$E_\alpha$ is the kinetic energy of $\alpha$ in laboratory frame,
and the number of the data is $N = 351$~\cite{tetal-prc09}.

\begin{table}[t]
\begin{center}
\begin{tabular}{c | c c c c} \hline
$\tilde g_0$  &  0 & 20 & 40 & 60  \cr \hline
$P_0$ (fm$^3$) & $-0.03573(3)$ & $-0.03575(3)$  & $-0.03578(3)$ & $-0.03584(3)$ 
 \cr
$Q_0$ (fm$^5$) & $0.002055(9)$ & $0.002894(9)$ & $0.005408(9)$ & $009598(9)$ 
 \cr
$R_2$ (fm$^2$) & $0.976(14)$ & $0.954(15)$ & $0.869(15)$ & $0.731(16)$ \cr
$E_r$ (MeV) & $4.88763(6)$ & $4.88763(6)$ & $4.88760(6)$ & $4.88758(6)$ \cr
$\Gamma_r$ (MeV) & $0.00168(4)$ & $0.00167(4)$ & $0.00164(4)$ & $0.00157(4)$
 \cr \hline
$|C_b|_0$ (fm$^{-1/2}$) & $41.0(1)$ & $30.9(0)$ & $20.2(0)$ & $14.2(0)$ \cr
$|C_b|_1$ (fm$^{-1/2}$) & $443(3)$ & $278(1)$ & $166(0)$ & $115(0)$ \cr \hline
\end{tabular}
\caption{
Values and errors of five parameters
$\{P_0, Q_0, R_2, E_r, \Gamma_r\}$ fitted to the phase shift $\delta_0$
of elastic $\alpha$-$^{12}$C scattering for $l=0$ 
using some values of $\tilde g_0$,
$\tilde g_0 = 0, 20, 40, 60$.
The ANCs, $|C_b|_0$ and $|C_b|_1$, for the $0_1^+$ and $0_2^+$ 
states of $^{16}$O are calculated by using each set of the fitted values 
of the parameters.
}
\label{table;parameters}
\end{center}
\end{table}
In Table~\ref{table;parameters}, 
fitted values and errors of the five parameters 
$\{P_0,Q_0,R_2,E_r,\Gamma_r\}$ using the four values of 
$\tilde g_0$, $\tilde g_0=0,20,40,60$ to the phase shift data $\delta_0$
are displayed. Values of the ANCs, $|C_b|_0$ and $|C_b|_1$, 
for the ground $0_1^+$ state and the first excited $0_2^+$ state
of $^{16}$O calculated by using each set of the fitted parameters 
are also displayed in the table. We obtain almost the same $\chi^2$ values 
for the four fittings, $\chi^2/N=1.60$, and one can see that 
the errors of those fitted parameters almost do not change for 
the four cases. One can see that the fitted values of $E_r$ 
and $\Gamma_r$ almost do not change either for all the $\tilde g_0$ values
and agree well to the experimental data within the error bars. 
A similar tendency can be seen for the fitted values of $P_0$ as well.
On the other hand, one can notice that a remarkable $\tilde g_0$ dependence
for the values of $Q_0$; $\tilde g_0$ and $Q_0$ are strongly correlated 
with each other, and $\tilde g_0$ and $Q-0$ cannot 
simultaneously be fitted by using the phase shift data.     
A minor but a significance $\tilde g_0$ dependence can be seen for the 
values of $R_2$. 
Because the self-energy contribution, $-2\kappa H_2(\tilde \eta)$, 
from the first excited $2_1^+$ state of $^{12}$C appears out of the $d$-wave 
coupling while $R_2$ accounts for the non-resonant shape 
of the phase shift data at high energies, $5.5< E_\alpha < 6.62$~MeV,
those contributions may become competitive at the high energies. 
\begin{figure}[t]
\begin{center}
  \includegraphics[width=12cm]{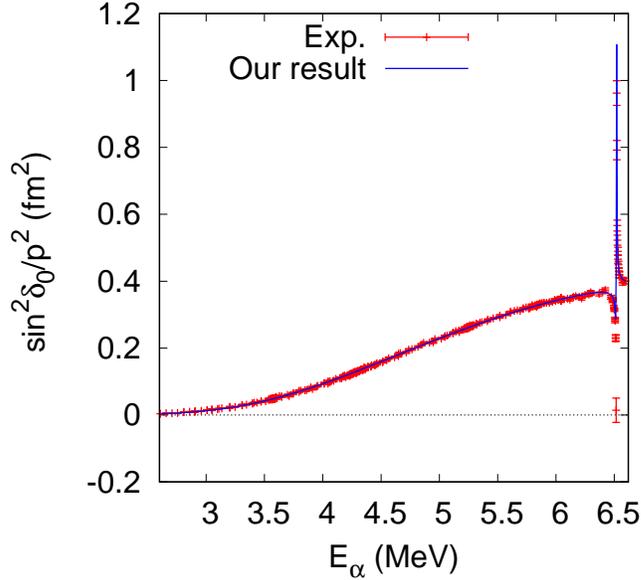}
\caption{
Squared amplitude, $\sin^2\delta_0/p^2$, as a function of 
$E_\alpha$ calculated by using the fitted values of the parameters in 
Table~\ref{table;parameters}.
Experimental data are included in the figure as well. 
}
\label{fig;A2}       
\end{center}
\end{figure}
In Fig.~\ref{fig;A2}, we plot a curve for the squared amplitude,
$\sin^2\delta_0/p^2$, as a function of $E_\alpha$ by using the fitted
values of the parameters; those four sets of the fitted parameters 
displayed in the table give almost the same curve plotted in the figure.   
We include the experimental data in the figure as well. One can see
the calculated curve well reproduces the experimental data. 

Regarding estimate of the ANCs, $|C_b|_0$ and $C_b|_1$,
for the ground $0_1^+$ state and the first excited $0_2^+$ state of
$^{16}$O, respectively, we find our results for the ANCs are significantly
sensitive to the $\tilde g_0$ values; those two ANCs decrease as the 
$\tilde g_0$ value increases while a value of $|C_b|_1$ is
about one order of magnitude larger than that of $|C_b|_0$ for a given
value of $\tilde g_0$. 
\begin{figure}[t]
\begin{center}
  \includegraphics[width=12cm]{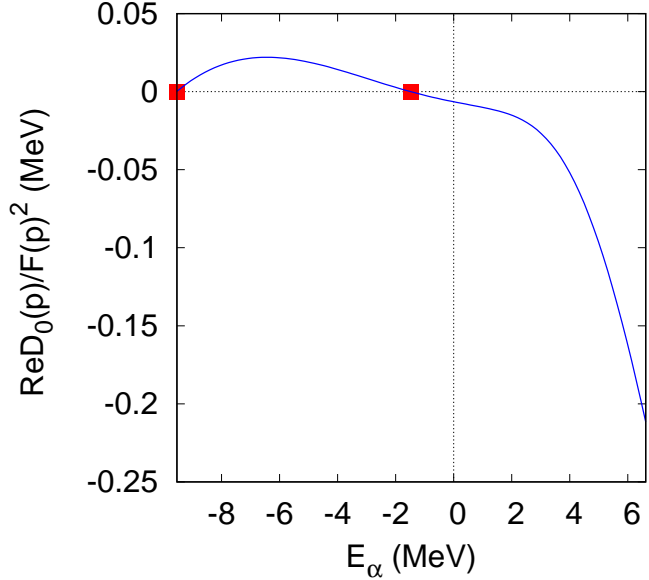}
\caption{
Real part of denominator of dressed $^{16}$O propagator for $l=0$, $D_0(p)$, 
including vertex form-factors $F(p)^2$ as a function of $E_\alpha$. 
Filled (red) squares denote the binding energies for the 
$0_1^+$ and $0_2^+$ states of $^{16}$O.
}
\label{fig;D0}       
\end{center}
\end{figure}
In Fig.~\ref{fig;D0}, we plot a curve for $D_0(p)/F(p)^2$ as a function
of $E_\alpha$; two filled (red) squares in the figure denote the binding 
energies of the two bound states. Those points are fixed 
in Eq.~(\ref{eq;binding_poles}), and the ANCs are calculated from the slope
of the curve at those points by using the relation shown in Eq.~(\ref{eq;Cb}). 
Because the function by which we plot the curve is given as a polynomial
function with the effective range parameters, the slope of the curve
becomes steep when the magnitude of $E_\alpha$ becomes large.
Thus, the slope at the ground $0_1^+$ state is steeper than that of the 
$0_2^+$ state, i.e., $|C_b|_0$ is smaller than $|C_b|_1$.   
In addition, when a value of $\tilde g_0$ becomes larger, the 
contribution from higher order terms in the polynomial function
(in the $-2\kappa H_2(\tilde \eta)$ function, compared to those
in the $-2\kappa H_0(\eta)$ term) become larger;
thus both the ANCs, $|C_b|_0$ and $|C_b|_0$, become smaller. 
Though we do not have any clue for a value of $\tilde g_0$,
we can fix it by using a value of one of the two ANCs, 
and then the other one can be predicted.    

We now discuss values of the ANCs found in previous studies and compare
them to our results. For the ANC, $|C_b|_1$, for the first excited $0_2^+$
state of $^{16}$O, we have $|C_b|_1 = 443 - 115$~fm$^{-1/2}$ for 
$\tilde g_0=0 - 60$. One can find in the literature~\cite{detal-rmp17} 
two groups for 
values of $|C_b|_1$: a large value group and a small value group.   
For the large value group, 
one may find three results, 
which are about more than 4 times larger than our result:
$|C_b|_1 = (15.6\pm 1.0)\times 10^2$~fm$^{-1/2}$ 
obtained from the $\alpha$ transfer reaction $^6$Li($^{12}$C,$d$)$^{16}$O
reported by Avila \textit{et al,}~\cite{aetal-prl15},
1800~fm$^{-1/2}$ from $R$-matrix analysis for broad level structure 
of $^{16}$O by deBoer \textit{et al.}~\cite{detal-prc13},   
and 1560~fm$^{-1/2}$ from $R$-matrix analysis for 
$^{12}$C($\alpha$,$\gamma$)$^{16}$O reaction 
by deBoer \textit{et al.}~\cite{detal-rmp17}.
For the small value group, one finds two results, 
which agree with our result:
$|C_b|_1 = 44^{+ 270}_{-44}$~fm$^{-1/2}$ from the study
of the $0_2^+$ state cascade transition 
in the $^{12}$C($\alpha$,$\gamma$)$^{16}$O reaction  
by Sch\"urmann \textit{et al.}~\cite{setal-plb11} and 
405.7~fm$^{-1/2}$ from so-called $\Delta$ method based on the 
effective range theory by Orlov, Irgaziev, and Nabi~\cite{oin-prc17}.

For the ANC, $|C_b|_0$, for the ground $0_1^+$ state of $^{16}$O, 
we have $|C_b|_0= 41.0 - 14.2$~fm$^{-1/2}$ for $\tilde g_0 = 0 - 60$.
One can also find two groups for values of $|C_b|_0$:
a large value group and a small value group in the literature.  
For the large value group, one may find two results, which are one
or two orders of magnitude larger than our result: 
709~fm$^{-1/2}$ from $E2$ interference effects 
in the $^{12}$C($\alpha$,$\gamma$)$^{16}$O reaction 
by Sayre \textit{et al.}~\cite{setal-prl12} and  
4000~fm$^{-1/2}$ (WS1), 1200~fm$^{-1/2}$ (WS2), and 750~fm$^{-1/2}$ (FP)
from a study of $^{12}$C($^{16}$O,$^{12}$C)$^{16}$O reaction,
where the results depend on the use of nuclear potentials:
Wood-Saxon 1 and 2 potentials (WS1, WS2) and folding potential (FP),
by Morais and Lichtenth\"aler~\cite{ml-npa11}.  
For the small value result, one may find four results, which agree
with our ones: $|C_b|_0 = 13.9(24)$~fm$^{-1/2}$ from a CDCC study 
for a resonant breakup of $^{16}$O by Adhikari and Basu~\cite{ab-plb09},
20.33~fm$^{-1/2}$ from the effective range expansion by Orlov, Irgaziev,
and Nikitina~\cite{oin-prc16}
and 21.76~fm$^{-1/2}$ from the $\Delta$ method based on
the effective range expansion by Orlov, Irgaziev, and Nabi~\cite{oin-prc17},
and 58~fm$^{-1/2}$ from the $R$-matrix analysis 
for the $^{12}$C($\alpha$,$\gamma$)$^{16}$O reaction 
by deBoer \textit{et al.}~\cite{detal-rmp17}.

\vskip 2mm \noindent
{\bf 5. Results and discussion}

In this work,
we studied an inclusion of a sharp resonant $0_3^+$ state of $^{16}$O
and a first excited $2_1^+$ state of $^{12}$C for the elastic $\alpha$-$^{12}$C 
scattering for $l=0$ channel up to the energy 
in which the sharp resonant $0_3^+$ state of $^{16}$O appears.
We separate the scattering amplitude into two parts;
one is an amplitude for the sharp resonant state, and the other is
for the rest of the non-resonant part of the amplitude.
The resonant part of the amplitude is presented as a Breit-Wigner-like
form while the non-resonant part of the amplitude is parameterized
by an effective range expansion.
A contribution from a bubble diagram due to the propagation of $\alpha$
and the $2_1^+$ state of $^{12}$C is included in the non-resonant part
of the amplitude. We also include a vertex correction for the initial
and final state interactions of $\alpha$ and $^{12}$C.
Four parameters appearing in the amplitude are fixed by using the 
binding energies for the $0_1^+$ and $0_2^+$ state and 
the resonant energy and the width for the $0_3^+$ states of $^{16}$O 
while the remaining four parameters,
$P_0$, $Q_0$, $\tilde g_0$, and $R_2$, are fitted to the experimental
phase shift data of the elastic $\alpha$-$^{12}$C scattering for $l=0$ channel.
We find a large correlation between $Q_0$ and $\tilde g_0$ and that 
a value of $\tilde g_0$, which represents a contribution from the 
$2_1^+$ state of $^{12}$C, is not fixed from the phase shift data.
While a vertex correction, $R_2$, for the initial and final states
of $\alpha$ and $^{12}$C is found to be crucial 
to reproduce the phase shift data
at an energy range, $5.5 \le E_\alpha \le 6.62$~MeV. 
We then calculate the ANCs for the $0_1^+$ and $0_2^+$ states of $^{16}$O.
We find that our numerical results of the ANCs significantly depend on 
a value of $\tilde g_0$ while values of the ANC for the $0_2^+$ are
about one order of magnitude are larger than those for the $0_1^+$
state. Our results of the ANCs are compared to those in the literature.
Scattered values of the ANCs in the previous results are found, and those
can be grouped into two, a large value group and a small value group 
for both the ANCs. Our results, we found, reasonably  well agree with 
those of the small value groups for both the ANCs.   

As one might have noticed, we did not mix 
the composite $^{16}$O fields, $d_{(rs)}$ and $d_{(nr)}$,
for the sharp resonant amplitude and the non-resonant
part of the amplitude.
Those two fields can be mixed in the amplitudes 
through the $\alpha$-$^{12}$C propagation;
in the $\alpha$-$^{12}$C bubble diagram, the $\alpha$-$^{12}$C state
is created through the $y_{(rs)}$ or $y_{(nr)}$ interaction, and,
after a propagation of $\alpha$ and $^{12}$C, they are destroyed
through the $y_{(rs)}$ or $y_{(nr)}$ interaction. 
Here we have assumed a naive counting rule in which the sharp resonant part 
of the amplitude becomes a leading order (LO) contribution near the 
resonant energy and at the off-resonant energy, the resonant part of the 
amplitude is suppressed and the non-resonant part of
the amplitude becomes a LO contribution. 
As a part of higher order corrections at next-to-next-to leading order
(NNLO), we rather phenomenologically included it as a vertex correction,
the $R_2$ term,
for the initial and final state interactions between $\alpha$ and $^{12}$C. 
We found that the correction is crucial to reproduce
the phase shift data at the energy region, $5.5\le E_\alpha \le 6.62$~MeV,
close to the sharp resonant energy $E_\alpha(0_3^+) = 6.52$~MeV. 
A complete treatment for the terms at NNLO would be interesting
for a future work. 

We also found a significant $\tilde g_0$ dependence in our numerical 
results for the ANCs for the $0_1^+$ and $0_2^+$ states of $^{16}$O
while a value of $\tilde g_0$ could not be fixed from the phase shift
data of the elastic $\alpha$-$^{12}$C scattering.
As mentioned above, a value of $\tilde g_0$ can be fixed 
by using an experimental datum of one of the two ANCs, 
and then we can predict the other one of the two ANCs,
though the values of the ANCs in the literature are
significantly scattered. 
Another way to fix $\tilde g_0$ is to use experimental data for
inelastic $\alpha$-$^{12}$C scattering, $\alpha + ^{12}$C($0_1^+$) 
$\to \alpha + {}^{12}$C$^*(2_1^+)$. 
For a better understanding of the present situation, further studies 
for the ANCs for the $0_1^+$ and $0_2^+$ states of $^{16}$O, 
both experimentally and theoretically, would be required

\vskip 2mm \noindent
{\bf Acknowledgements}

The author would like to thank A. Hosaka and T. Sato
for useful discussions.
This work was supported by Research Grant of Sunmoon University 2019.


\vskip 3mm \noindent
\begin{thebibliography}{99}


\bibitem{f-rmp84}
W.~A. Fowler,
Rev. Mod. Phys. \textbf{56}, 149 (1984).

\bibitem{lt-rmp58}
A.~M. Lane and R.~G. Thomas,
Rev. Mod. Phys. {\bf 30}, 257 (1958).

\bibitem{bb-npa06}
L.~R. Buchmann and C.~A. Barnes,
Nucl. Phys. A {\bf 777}, 254 (2006).

\bibitem{chk-epja15}
A. Coc, F. Hammache, J. Kiener,
Eur. Phys. J. A {\bf 51}, 34 (2015).

\bibitem{bk-ppnp16}
C.~A. Bertulani and T. Kajino,
Prog. Part. Nucl. Phys. {\bf 89}, 56 (2016).

\bibitem{detal-rmp17}
R.~J. deBoer {\it et al.},
Rev. Mod. Phys. {\bf 89}, 035007 (2017),
and references therein.

\bibitem{a-epja16}
S.-I. Ando,
Eur. Phys. J. A \textbf{52}, 130 (2016).

\bibitem{a-prc18}
S.-I. Ando,
Phys. Rev. C \textbf{97}, 014604 (2018).

\bibitem{a-jkps18}
S.-I. Ando,
J. Korean Phys. Soci. \textbf{73}, 1452 (2018).

\bibitem{ya-jkps19}
H.-E. Yoon and S.-I. Ando,
J. Korean Phys. Soci. \textbf{75}, 202 (2019).

\bibitem{a-prc19}
S.-I. Ando,
Phys. Rev. C \textbf{100}, 015807 (2019).

\bibitem{a-19}
S.-I. Ando, work submitted to Eur. Phys. J. A.

\bibitem{bv-arnps02}
P.F. Bedaque and U. van Kolck,
Ann. Rev. Nucl. Part. Sci. {\bf 52}, 339 (2002).

\bibitem{bh-pr06}
E. Braaten and H.-W. Hammer,
Phys. Rept. {\bf 428}, 259 (2006).

\bibitem{m-15}
U.-G. Mei\ss ner,
Phys. Scripta {\bf 91}, 033005 (2016).

\bibitem{hjp-17}
H.-W. Hammer, C. Ji, D.R. Phillips,
J. Phys. G {\bf 44}, 103002 (2017).

\bibitem{hkvk-19}
H.-W. Hammer, S. Konig, and U. van Kolck,
Rev. Mod. Phys. \textbf{92}, 25004 (2020).

\bibitem{hhvk-npa08}
R. Higa, H.-W. Hammer, and U. van Kolck,
Nucl. Phys. A \textbf{809}, 171 (2008).

\bibitem{g-prc09}
B.~A. Gelman,
Phys. Rev. C \textbf{80}, 034005 (2009).

\bibitem{tetal-prc09}
P. Tischhauser \textit{et al.},
Phys. Rev. C \textbf{79}, 055803 (2009).

\bibitem{a-epja07}
S.-I. Ando,
Eur. Phys. J. A \textbf{33}, 185 (2007).

\bibitem{b-pr49}
H.~A. Bethe,
Phys. Rev. {\bf 76}, 38 (1949).

\bibitem{k-npb97}
D.~B. Kaplan,
Nucl. Phys. B {\bf 494}, 471 (1997).

\bibitem{bs-npa01}
S.~R. Beane and M.~J. Savage,
Nucl. Phys. A {\bf 694}, 511 (2001).

\bibitem{ah-prc05}
S. Ando and C.~H. Hyun,
Phys. Rev. C {\bf 72}, 014008 (2005).

\bibitem{ir-prc84}
Z.~R. Iwinski and L. Rosenberg,
Phys. Rev. C \textbf{29}, 349 (1984).

\bibitem{twc-npa93}
D.~R. Tilley, H.~R. Weller, and C.~M. Cheves,
Nucl. Phys. A \textbf{564}, 1 (1993).

\bibitem{fmetal-12}
D. Foreman-Mackey {\it et al.},
Publ. Astron. Soc. Pac. \textbf{125}, 306 (2013). 

\bibitem{aetal-prl15}
M.~L. Avila \textit{eta al.},
Phys. Rev. Lett. \textbf{114}, 071101 (2015).

\bibitem{detal-prc13}
R.~J. deBoer \textit{et al.},
Phys. Rev. C \textbf{87}, 015802 (2013).

\bibitem{setal-plb11}
D. Sch\"urmann \textit{et al.},
Phys. Lett. B \textbf{703}, 557 (2011).

\bibitem{oin-prc17}
Yu.~V. Orlov, B.~F. Irgaziev, and J.-U. Nabi,
Phys. Rev. C \textbf{96}, 025809 (2017). 

\bibitem{setal-prl12}
D.~B. Sayre \textit{et al.},
Phys. Rev. Lett. \textbf{109}, 142501 (2012).

\bibitem{ml-npa11}
M.~C. Morais and R. Lichtenth\"aler,
Nucl. Phys. A \textbf{857}, 1 (2011).

\bibitem{ab-plb09}
S. Adhikari and C. Basu,
Phys. Lett. B \textbf{682}, 216 (2009).

\bibitem{oin-prc16}
Yu.~V. Orlov, B.~F. Irgaziev, and L.~I. Nikitina,
Phys. Rev. C \textbf{93}, 014612 (2016).


\end{thebibliography}
\end{document}